\documentclass{ws-ijbc}
\usepackage{ws-rotating}     
\usepackage{color}
\usepackage{graphicx}
\usepackage{epstopdf}
\usepackage{lineno}
\begin{document}

\catchline{}{}{}{}{} 

\markboth{Scarsoglio et al.}{Complex networks unveiling spatial patterns in turbulence}

\title{COMPLEX NETWORKS UNVEILING SPATIAL PATTERNS IN TURBULENCE}

\author{STEFANIA SCARSOGLIO}
\address{Department of Mechanical and Aerospace Engineering, Politecnico di Torino\\
Corso Duca degli Abruzzi 24, Torino, Italy\\
stefania.scarsoglio@polito.it}

\author{GIOVANNI IACOBELLO}
\address{Department of Mechanical and Aerospace Engineering, Politecnico di Torino\\
Corso Duca degli Abruzzi 24, Torino, Italy\\
giovanni.iacobello@studenti.polito.it}

\author{LUCA RIDOLFI}
\address{Department of Environmental, Land and Infrastructure Engineering, Politecnico di Torino\\
Corso Duca degli Abruzzi 24, Torino, Italy\\
luca.ridolfi@polito.it}

\maketitle

\begin{history}
\received{May 4, 2016; Revised July 18, 2016}
\end{history}


\begin{abstract}
Numerical and experimental turbulence simulations are nowadays reaching the size of the so-called \emph{big data}, thus requiring refined investigative tools for appropriate statistical analyses and data mining. We present a new approach based on the complex network theory, offering a powerful framework to explore complex systems with a huge number of interacting elements. Although interest on complex networks has been increasing in the last years, few recent studies have been applied to turbulence. We propose an investigation starting from a two-point correlation for the kinetic energy of a forced isotropic field numerically solved. Among all the metrics analyzed, the degree centrality is the most significant, suggesting the formation of spatial patterns which coherently move with similar vorticity over the large eddy turnover time scale. Pattern size can be quantified through a newly-introduced parameter (i.e., average physical distance) and varies from small to intermediate scales. The network analysis allows a systematic identification of different spatial regions, providing new insights into the spatial characterization of turbulent flows. Based on present findings, the application to highly inhomogeneous flows seems promising and deserves additional future investigation.
\end{abstract}

\keywords{complex networks; turbulent flows; time series analysis; spatial correlation; spatiotemporal patterns}

\section{Introduction}

Turbulence is an important and widely investigated topic, involving everyday life in several natural phenomena (e.g., rivers, bird flight and fish locomotion, atmospheric and oceanic currents) and industrial applications (e.g., flow through pumps, turbines, chemical reactors, and aircraft-wing tips). Although studied for decades \cite{Frisch}, due to its chaotic and complex nature, several important questions regarding its spatial characterization, prediction, and control remain mostly unclear \cite{Warhaft}. In order to achieve a better description of its dynamic, nowadays experimental and numerical simulations progressively provide a greater amount of extremely detailed data, which need to be examined and interpreted. There is therefore an increasing urgency of refined investigative tools for appropriate statistical analyses and data mining. Different and interdisciplinary approaches, borrowed from bioinformatics to physical statistics, can help exploring data from a complementary and innovative perspective.

In the last years, interest in complex network theory has grown enormously, as it offers a synthetic and powerful tool to study complex systems with an elevated number of interacting elements \cite{albert_barabasi_2002,WS1998,newman2010}. By combining graph theory and statistical physics, the present approach find immediate applications to real existing networks (e.g., Word Wide Web, social, economical and neural connections) as well as in building networks from spatio-temporal data series \cite{costa_et_al_2011,Boccaletti2006}. A relevant example is represented by the climate networks, where different meteorological series have been transformed into networks to disentangle the global atmospheric dynamics (see, among others, \cite{Yamasaki_et_al_2008,steinhaeuser_et_al_2012,scarsoglio2013,sivakumar2014,tsonis_et_al_2008,Donges_et_al_2009}).

In turbulence, few and very recent network-based approaches have been proposed to characterize patterns in two-phase stratified flows \cite{gaojin2009,Gao2013,Gao2015exp,gao2015,Gao2015SciRep,gao2015epl,Gao2016}, turbulent jets \cite{shirazi,Charakopoulos}, as well as reacting \cite{Murugesan} and fully developed turbulent flows \cite{Liu,manshour}. Most of them focused on temporal data measured in different spatial locations and, by means of the visibility algorithm \cite{Lacasa} or recurrence plots \cite{Donner,Marwan}, converted each time series into a network. Because of the promising results so far obtained and the potentiality of the network tools, turbulence networks certainly merit further investigation.

We here proposed a complex network analysis on a forced isotropic turbulent field solved through direct numerical simulation (DNS), available from the Johns Hopkins Turbulence Database (JHTDB) \cite{JHTDB1,JHTDB2}. Differently to what was carried out so far, we did not transform each temporal series into a network but constructed a single global network from spatio-temporal data. The network was built starting from a two-point correlation for the turbulent kinetic energy computed over all the couples of the selected nodes. In so doing, a unique monolayer network was obtained, whose nodes partially overlap the numerical grid cells and whose links are active if the distance and statistical interdependence between two nodes satisfy suitably chosen constraints \cite{Donges_et_al_2009}.
Correlation-based networks \cite{Donner,Yang} is probably the most used way of applying network science techniques to time series, with examples ranging from financial markets \cite{Caraiani} to brain activity \cite{Stam}. However, to the best of our knowledge, the application of correlation networks to spatio-temporal turbulent data has not been analyzed to date.

\noindent Once the network was built, different topological features were analyzed. The degree centrality turned out to be the most meaningful parameter, suggesting the onset and evolution of spatial patterns which coherently move with similar vorticity over the large eddy turnover time scale. A new network metric here introduced (i.e., average physical distance) is able to indicate the spatial scale of the turbulent patterns, ranging from small to intermediate scales.

\section{Methods}

\subsection{Johns Hopkins Turbulence Database Description}

The forced isotropic turbulence field here used was solved by means of a DNS over $1024^3$ nodes and is available from the JHTDB \cite{JHTDB1,JHTDB2}. Velocity ($u, v, w$), vorticity ($\omega_x, \omega_y, \omega_z$), and pressure ($p$) fields were computed over a cube of dimension $2 \pi$ x $2 \pi$ x $2 \pi$. A forcing term was added to the Navier-Stokes equations so that the total kinetic energy does not decay and, after a transient range, the field can be considered statistically stationary. Once this state was reached, 1024 frames of data were recorded (time-step=0.002), lasting about one large-eddy turnover time, $T_L$. Energy was injected by keeping the total energy constant, so that only the integral scale is influenced by the forcing, while the intermediate and the dissipative ranges are not involved. Some statistical characteristics are here given together with a brief physical recall:

\begin{itemlist}
\item Taylor microscale, $\lambda = 0.118$. The Taylor microscale is the intermediate turbulent length scale, between the integral and the Kolmogorov scales, at which turbulent eddies are still substantially influenced by viscosity;
\item Taylor-scale Reynolds number, $Re_{\lambda} = (u_{rms} \lambda)/\nu = 433$, is the ratio between inertial and viscous forces at the Taylor scale, $\lambda$ ($u_{rms}$ is the root-mean-square velocity and $\nu$ is the kinematic viscosity);
\item Kolmogorov time scale, $\tau_k = 0.0446$, and length scale, $\eta = 0.00287$. These are the smallest scales in turbulence, where viscosity dominates and the turbulent kinetic energy is dissipated;
\item integral scale, $L = 1.376$, is the size of the largest eddies of the flow;
\item large eddy turnover time, $T_L=2.02$, is the time scale over which the largest eddies develop.
\end{itemlist}

\noindent The JHTDB provides an accurate multi-terabyte and comprehensive data archive, which has been widely exploited for testing modeling \cite{LES}, structural properties \cite{Lawson}, experimental data \cite{Exper} and statistical analyses \cite{Kolm}.

\subsection{Complex network metrics}

The network measures used in the present work are here summarized \cite{albert_barabasi_2002,Boccaletti2006}. A network is defined by a set $V = {1, ..., N}$ of nodes and a set $E$ of links $\{i,j\}$. We assume that a single link can exist between a pair of nodes. The \emph{adjacency matrix}, $A$:

\begin{equation}
\label{adjacency} A_{ij} = \begin{cases} 0, & \mbox{if } \{i,j\} \notin E\\ 1, & \mbox{if } \{i,j\} \in E, \end{cases}
\end{equation}

\noindent accounts whether a link is active or not between nodes $i$ and $j$. The network is considered as undirected, thus $A$ is symmetric, and no self-loops are allowed ($A_{ii}=0$).

\noindent The \emph{normalized degree centrality} of a node $i$ is defined as

\begin{equation}
\label{dc} k_i = \frac{\sum\limits_{j=1}^N A_{ij}}{N-1},
\end{equation}

\noindent and gives the number of neighbors of the node $i$, normalized over the total number of possible neighbors ($N-1$). We also define $K_i = k_i (N-1)$ as the (non-normalized) degree centrality.

\noindent The \emph{eigenvector centrality}, measuring the influence of the node $i$ in the network, is given by

\begin{equation}
x_i = \frac{1}{\lambda} \sum_k A_{ki} x_k,
\end{equation}

\noindent with $A_{ki}$ the adjacency matrix and $\lambda$ its largest eigenvalue \cite{newman2010}. In matrix notation, we can write:

\begin{equation}
{\lambda} x = x A,
\end{equation}

\noindent where the centrality vector $x$ is the left-hand eigenvector of the adjacency matrix $A$ associated with the eigenvalue $\lambda$, which is the largest eigenvalue in absolute value.

\noindent The \emph{local clustering coefficient} of a node is

\begin{equation}
C_i = \frac{e(\Gamma_i)}{\frac{K_i(K_i-1)}{2}},
\end{equation}

\noindent where $e(\Gamma_i)$ is the number of links connecting the vertices within the neighborhood $\Gamma_i$, and $K_i(K_i-1)/2$ is the maximum number of edges in $\Gamma_i$. The local clustering coefficient represents the probability that two randomly chosen neighbors of a node are also neighbors.

\noindent The \emph{betweenness centrality} of a node is

\begin{equation}
BC_k = \sum_{i,j \neq k} \frac{\sigma_{ij}(k)}{\sigma_{ij}},
\end{equation}

\noindent where $\sigma_{ij}$ are the number of shortest paths connecting nodes $i$ and $j$, while $\sigma_{ij}(k)$ represents the number of shortest paths from $i$ to $j$ through node $k$. If node $k$ is crossed by a large number of all existing shortest paths (i.e. if $BC_k$ is large), then node $k$ is reputed an important mediator for the information transport in the network.

\noindent \emph{Modularity} $Q$ is a measure of the structure of networks, detecting the presence of communities/modules \cite{newman2004}. $Q$ is defined, up to a multiplicative constant, as the fraction of the edges that fall within the given groups minus the expected such fraction if edges were distributed at random. A high modularity degree (roughly above 0.3) indicates a strong division of the network into clusters \cite{newman2006}. $Q$ can be mathematically quantified as

\begin{equation}
Q = \frac{1}{4 m} \sum_{ij} \left(A_{ij} - \frac{K_i K_j}{2 m} \right) s_i s_j,
\end{equation}

\noindent where $A_{ij}$ is the adjacency matrix, $(K_i K_j)/(2 m)$ is the expected number of edges between nodes $i$ and $j$ if edges are placed at random, $m$ is the total number of links in the network, $s$ is a membership variable considering that the graph can be partitioned into two communities ($s_i=1$ if node $i$ belongs to community 1, $s_i=-1$ if it belongs to community 2), while $1/(4m)$ is merely conventional.

\noindent In the end, we introduce a new metric which is related to the reciprocal physical distance of the network nodes. The \emph{neighborhood physical distance}, $L_i$, of a node $i$ is the averaged physical distance between node $i$ and its neighborhood $\Gamma_i$:

\begin{equation}
L_i = \frac{\sum\limits_{j \in \Gamma_i} l_{ij}}{K_i},
\end{equation}

\noindent where $l_{ij}$ is the physical distance between node i and its neighbor $j$, $K_i$ is the degree centrality of node $i$.

\subsection{Building the network}

To build the network, we considered a spherical subdomain with center $C=(391,391,512)$ and radius $r=0.24$. For all the nodes inside this sphere we computed the kinetic energy time series, $E =(u^2 +v^2 + w^2)/2$. This local scalar variable is directly based on the primary flow field variables and is crucial to characterize the turbulent network. In fact, starting from the energy time series at a fixed point, we can infer what happens in its spatial surroundings. This choice allowed us to define a monolayer network, by evaluating the temporal linear correlation among all the cells of the sphere through the correlation matrix, $R_{ij}$. A linear Pearson correlation was adopted, as it is one of the simplest possible metrics to quantify the level of statistical interdependence between the temporal series. To avoid results biased by the network geometry, a link between nodes $i$ and $j$ exists if the following conditions are simultaneously satisfied:

\begin{itemlist}
\item $|R_{ij}|>\tau$, where $\tau$ is a suitable threshold;
\item At least one between nodes $i$ and $j$ lies inside the reference sphere with radius $r=0.12$ and center $C$;
\item The physical distance between nodes $i$ and $j$ is less or equal to $r=0.12$.
\end{itemlist}

In so doing, every node $i$ within the reference sphere had a well-defined region of influence (a sphere with radius 0.12 and centered in the node $i$ itself) where links with other nodes can occur. The region of influence had the same size for all nodes, so that every node within the reference sphere experienced the same number of potential links.

\noindent The size of the reference sphere ($r=0.12$) is linked to the Taylor scale, $\lambda=0.118$, as we are interested in what happens at scales of this order or smaller, where the spatial correlation is high, therefore avoiding spurious correlations which may occur at larger distances. Recall that in isotropic turbulence, by statistically averaging over an adequate number of samples, the two-point correlation function smoothly goes to zero as distance increases. Long range links, if present, can only be consequence of spotted random correlations, which may disturb real short-range links and spuriously alter the network metrics. For this reason, we restrict the maximum link length to the region ($r \leq \lambda$) where the noisy links are not present. The turbulent field is isotropic as a consequence of the DNS geometry and boundary conditions imposed, thus no preferential directions can be detected. Moreover, since the forcing to keep the total energy constant acts on bigger scales (wavenumber $|w| \leq 2$), intermediate and small scales do not experience any source of inhomogeneity. Therefore, the center of the reference sphere can be arbitrarily chosen and we selected the point $C$. To test the sensitivity of the results, another domain portion was then analyzed, namely a spherical subdomain with radius $r=0.24$ centered in $C'=(530,673,475)$.

The selection of the threshold, $\tau$, was a non-trivial aspect of building the network and had to take into account the goal of both evidencing strong spatial correlations and managing an appropriate number of nodes. The influence of the threshold has been deeply analyzed in climate networks \cite{Donges_et_al_2009}. The threshold $\tau=0.9$ represents a good compromise between a very high degree of correlation and a suitable network cardinality. A sensitivity analysis on $\tau$ values is reported in the Results and Discussion section.

The network is composed by $N=128785$ nodes and $m=80920781$ links, indicating with $N_{int}=31343$ the cardinality of nodes inside the reference sphere and with $N_{ext}$ the number of nodes outside of it ($N=N_{int} + N_{ext}$). The edge density, $\rho(\tau)$, is defined as

\begin{equation}
\rho(\tau) = \frac{n(\tau)}{\frac{N(N-1)}{2} - \frac{N_{ext}(N_{ext}-1)}{2}},
\end{equation}

\noindent where $n(\tau)$ is the number of active links when the absolute value of $R_{ij}$ is above the threshold $\tau$ for the two-point correlation. The denominator accounts for the total number of links of the network, excluding links between purely external nodes (links between internal and external nodes are allowed). The edge density, $\rho$, is the ratio between active links above a given threshold $\tau$ and the total number of possible links. For the chosen threshold, $\rho=2.282 \cdot 10^{-2}$. The combined bidimensional edge density, $\rho_{2}(\tau,l)$, is introduced as

\begin{equation}
\rho_{2}(\tau,l) = \frac{n(\tau,l)}{\textmd{\# total links with distance \emph{l}}},
\end{equation}

\noindent where $l \in (0, 0.12]$ is the physical distance between two nodes, $n(\tau,l)$ is the number of active links above the threshold $\tau$ and at a fixed $l$. The combined bidimensional edge density, $\rho_2$, is the ratio of active links above a given threshold $\tau$ at a fixed distance $l$ and the total number of potential links at the same distance $l$. A graphical representation of $\rho_{2}(\tau,l)$ is reported in Fig. 1, where high density values are found for small physical distances, confirming that at $\tau=0.9$ short-term links are always active ($\rho \rightarrow 1$ if $l \rightarrow 0$). It should be noted that the combined bidimensional edge density at a fixed $\tau$ represents the link length distribution. To summarize, $\rho$ evaluates the density of active links independently of their physical lengths, while $\rho_2$ is the link density as function of the length.

\begin{figure}[h]
\centering
\includegraphics[width=0.6\columnwidth]{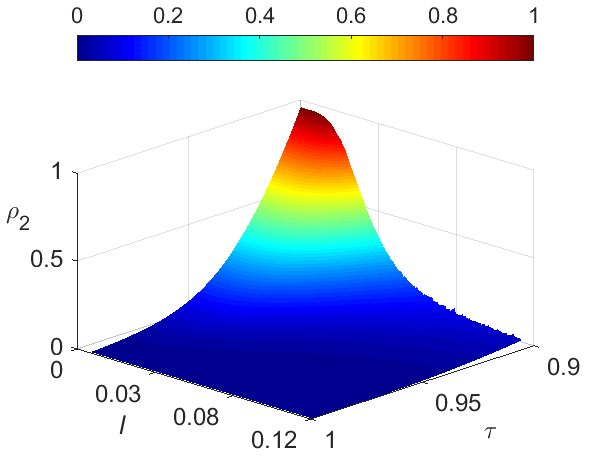}
\caption{Combined bidimensional edge density, $\rho_{2}(\tau,l)$.}
\label{edge}
\end{figure}

\noindent The network analysis presented in the following section is focused on the set of internal nodes, $N_{int}$, of the reference sphere. External nodes, which are part of the network but only exploited to evaluate links between internal and external nodes, are not shown.

\section{Results and Discussion}

\begin{figure}[h!]
\includegraphics[width=\columnwidth]{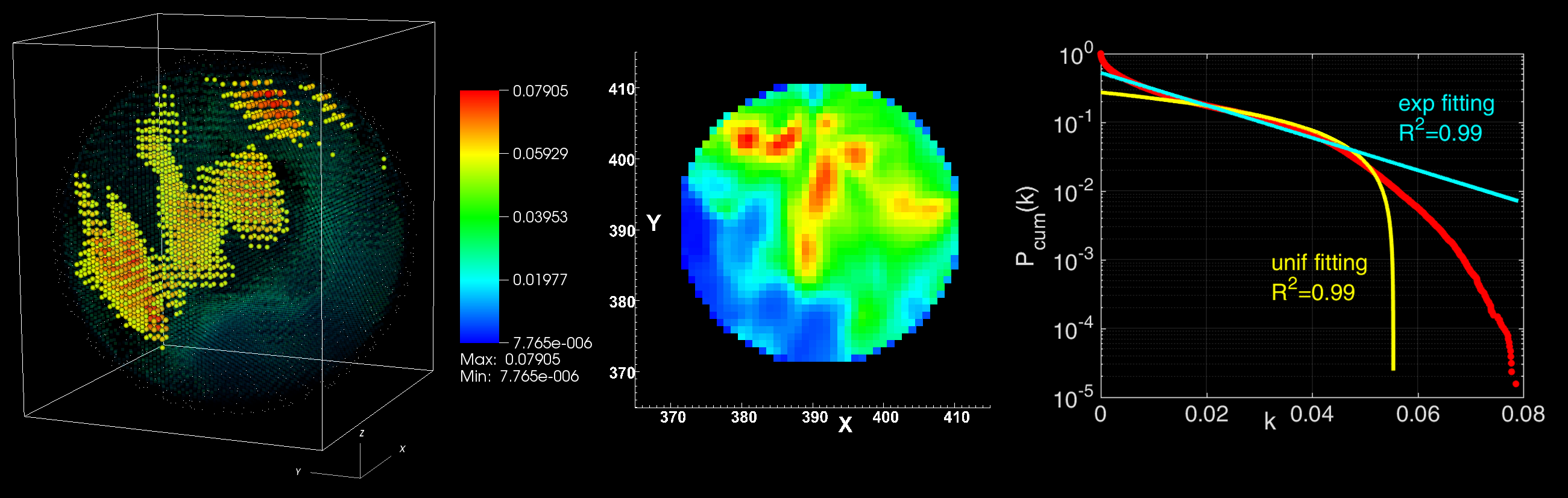}
\caption{Normalized degree centrality, $k$. (left) 3D perspective (values higher than 70 $\%$ of the maximum value are reported with points, the the rest of the network is transparently colored). (center) 2D section on the $z=512$ plane. (right) Cumulative degree distribution function, $P_{cum}(k)$, with the exponential and uniform fittings of the data and the corresponding coefficients of determination, $R^2$. A semi-log graph is adopted.}
\label{DC}
\end{figure}

\begin{figure}[h!]
\includegraphics[width=\columnwidth]{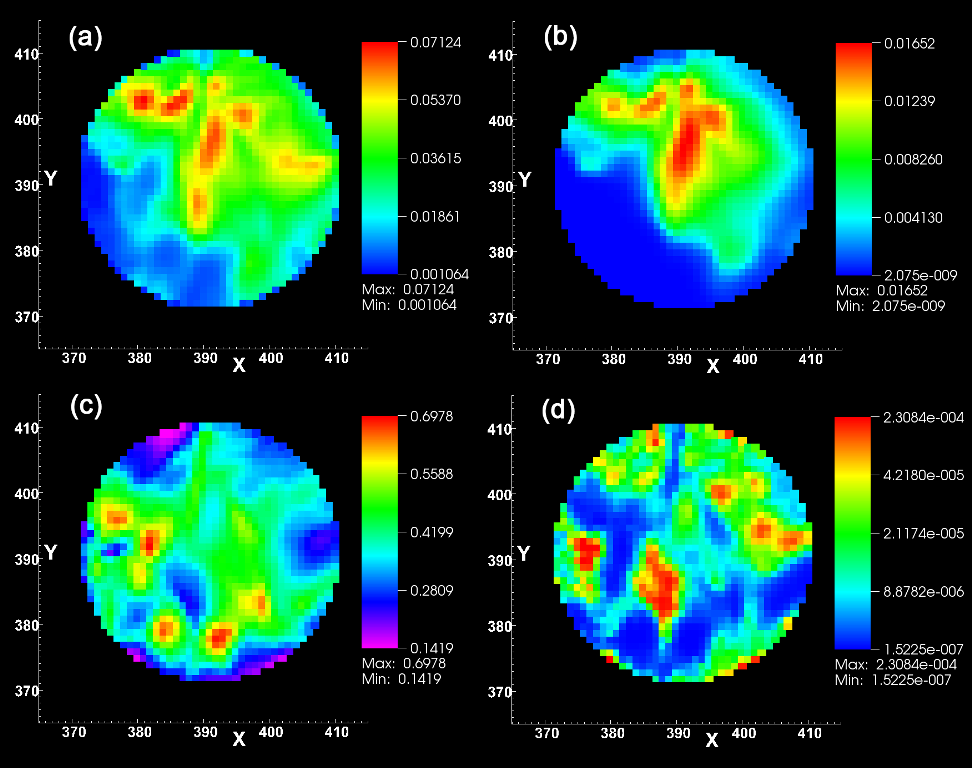}
\caption{(a) Normalized degree centrality, (b) eigenvector centrality, (c) local clustering coefficient, (d) betweenness centrality (a non-linear color scale is used). Results are displayed on the section $z=512$.}
\label{comparison}
\end{figure}

The properties of the turbulence network are here discussed. The degree centrality was first analyzed, evidencing regions with high values which are clearly distinguishable from the rest of the network. In Fig. 2 (left panel) the highest values (above 70 $\%$ of the maximum value) are highlighted through a 3D perspective, while the other values are transparently colored. A 2D diametral section on the plane $z=512$ is also displayed, reporting all $k_i$ values (central panel). In the right panel, the cumulative degree distribution function, $P_{cum}(k)=\sum_{k'=k}^{\infty} p(k')$, is shown in a linear-log plot. The degree distribution is adequately fitted by an exponential distribution for low $k$ values ($k<0.05$), as happens in many real world complex networks \cite{Dunne,Deng}. The right-tail has a qualitative downward behavior \cite{Dunne}, with a decay which is faster than an exponential but slower than an uniform distribution. Moreover, the network presents a rich-club effect \cite{Boccaletti2006}, i.e. high degree vertices connect one to each other.

Other network properties, such as the eigenvector centrality, the local clustering coefficient and the betweenness centrality, are reported and compared with the degree centrality in Fig. 3, as sections of the plane $z=512$. The eigenvector centrality (panel b) carried the same information of the degree centrality (a), confirming the presence of distinct spatial regions with high correlation. To this end, it should be noted that a completely random flow field would result in a highly disconnected network, which in turn would entail a spotted distribution for the centrality indexes, with the most part of values close to zero. The local clustering coefficient (panel c) was poorly related to the degree centrality, as in general happens in spatial networks \cite{Boccaletti2006}. The betweenness centrality (panel d) presented quite low values and a spotted distribution over the section, which weakly correlates to the degree centrality. No sources of inhomogeneity and anisotropy were present in the field, thus there are no preferential pathways transporting the information. This translated into a spotted distribution, with significantly high $BC$ gradient values, which is scarcely informative from the point of view of pattern formation.

The modularity value of the present network is about $Q=0.31$, and twenty-eight communities were detected through the Newman algorithm \cite{newman2004,newman2006}, where $Q=\sum_{c=1}^{28} q(c)$ and $q(c)$ is the modularity of a single community. Modularity is not uniformly distributed over the communities (Fig. 4, left panel), as the last eight modules have $q$ values close to zero, while the first community has the highest $q$ value (0.055), which is about $18\%$ of the total modularity value, $Q$. Nodes belonging to the community with the highest modularity $q$ are reported in Fig. 4 (right panel, red points). This community is the largest in terms of cardinality and detects a cluster of nodes which are physically close one to each other, representing a wide coherent region sharing the same properties. Moreover, high degree centrality values are usually found for nodes belonging to high-order communities. In particular, about $86\%$ of the nodes highlighted in Fig. 2 (left panel) falls within the first eight communities. The latest communities are instead less populated with nodes having medium to low degree centrality values. From the present findings, nodes seem to be partitioned into communities based on their reciprocal physical distance and on their connection to high centrality nodes.

\begin{figure}[h]
\includegraphics[width=\columnwidth]{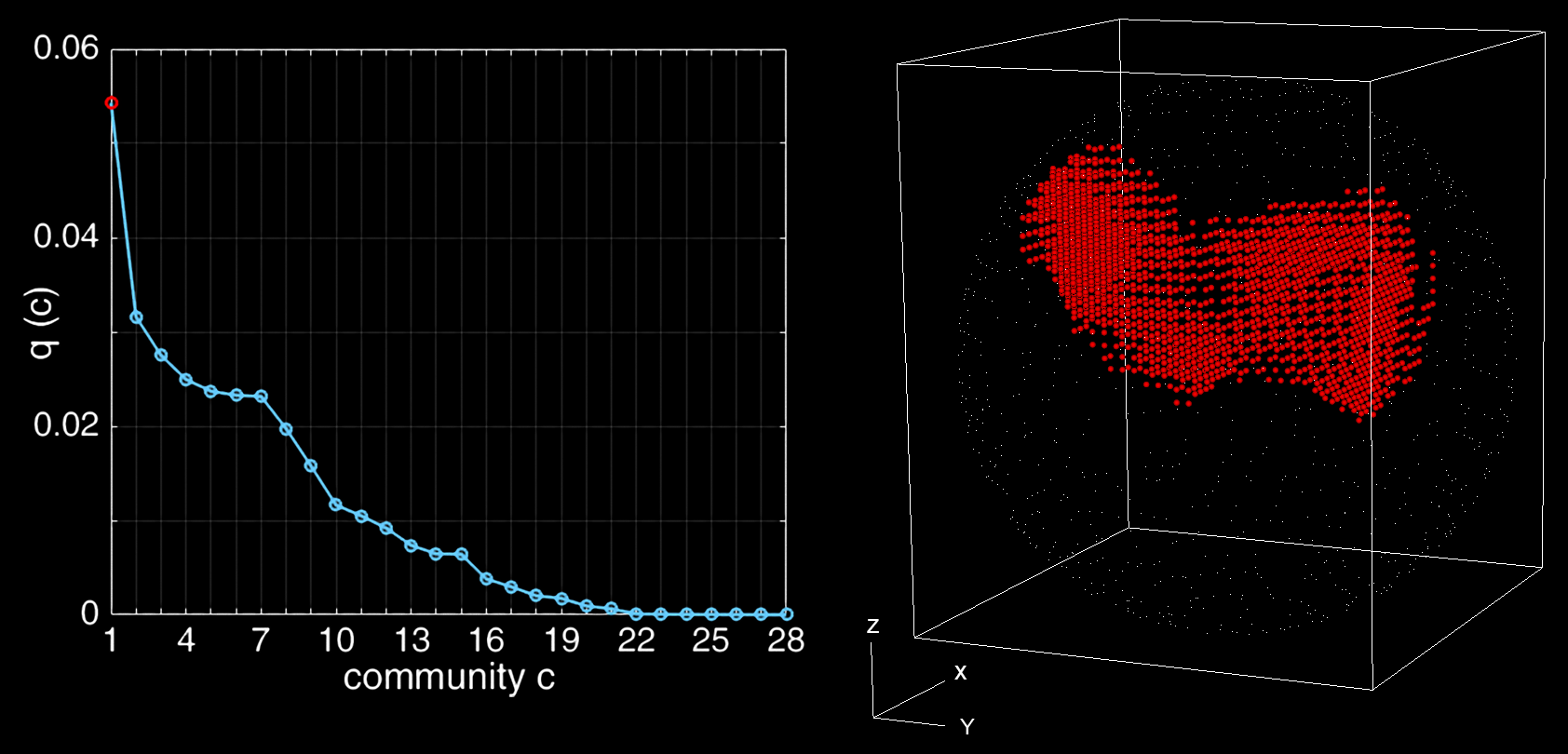}
\caption{(left) Modularity distribution over the twenty-eight communities, $q(c)$. (right) Nodes belonging to the community with the highest modularity value ($q=0.055$).}
\label{modularity}
\end{figure}

The most meaningful parameter turned out to be the degree centrality together with the eigenvector centrality, both direct measures of the importance of a node in the network. In order to interpret the network results in terms of physical properties of the turbulent field, we considered the highest degree centrality node (node HDC, $k=7.905 \cdot 10^{-2}$, coordinates (385,401,508)) and another with very low degree centrality (node LDC, $k=3.052 \cdot 10^{-3}$, coordinates (372,387,510)). For both nodes we evaluated their neighborhoods ($\Gamma(HDC)=10180$ and $\Gamma(LDC)=393$) and the average physical distance ($L_{HDC}=7.515 \cdot 10^{-2}$, $L_{LDC}=2.985 \cdot 10^{-2}$). In Fig. 5 (left) HDC and LDC nodes are shown together with the respective neighborhoods. We then considered nodes A and B at an intermediate physical distance $6.142 \cdot 10^{-2}$ (10 grid cells) from nodes HDC and LDC, respectively. Nodes A and B have normalized degree centrality values $k=4.057 \cdot 10^{-2}$ and $k=1.553 \cdot 10^{-5}$, respectively. We evaluated the temporal series of the vorticity modulus ($|\omega|= \sqrt{\omega_x^2 + \omega_y^2 + \omega_z^2}$) for the two pairs of nodes, (HDC-A) and (LDC-B). The couple (HDC-A) presented a strong temporal correlation for $|\omega|$ ($R=0.92$) and the two time series showed values close one to the other. The couple (LDC-B) had a much weaker correlation for $|\omega|$ ($R=0.68$) and the two time series often reached very different specific values (Fig. 5, right). The behaviour of the pairs (HDC-A) and (LDC-B) is representative of high degree centrality and low degree centrality regions, since analogous comparisons were found for many other couples of nodes. In Table 1 examples of couples of nodes showing the mentioned behaviours are shown. Thus, we can say that high degree centrality values indicate regions with the same instantaneous vorticity, that is turbulent patterns coherently moving over the time scale $T_L$. Moreover, there is a direct correlation between the degree centrality, $k$, and the average physical distance, $L$, of a node. $L$ gives the order of magnitude of the spatial patterns identified by the $k_i$ distribution. For node HDC, $L=7.515 \cdot 10^{-2}$, for node LDC, $L=2.985 \cdot 10^{-2}$, meaning that the size of the patterns ranges between the dissipative scale and the Taylor microscale.

\begin{figure}[h]
\includegraphics[width=\columnwidth]{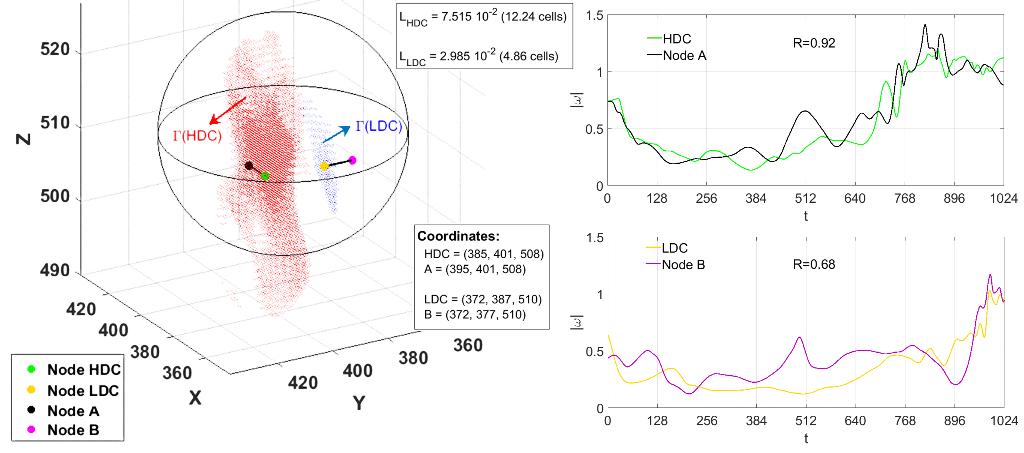}
\caption{(left) High degree centrality (HDC) and low degree centrality (LDC) nodes shown together with their neighborhoods ($\Gamma(HDC)$ in red, $\Gamma(LDC)$ in blue). Nodes A and B are at a distance $6.142 \cdot 10^{-2}$ from nodes HDC and LDC, respectively ($l_{HDC,A}=l_{LDC,B}=6.142 \cdot 10^{-2}$, 10 grid cells). (right) Time series of the vorticity modulus $|\omega|$ are shown for the pairs (A-HDC) and (B-LDC) with the corresponding correlation coefficient, $R$.}
\label{discussion}
\end{figure}

\begin{table}[h]
\tbl{Examples of couples of nodes belonging to high and low degree centrality regions.}
{\begin{tabular}{|c|c|c|c|}
  \hline
  & Nodes & Distance & R \\
  \hline
  High & $HDC_1$=(382,405,515), $A_1$=(382,405,507) & 8 grid cells & 0.96 \\
  \cline{2-4}
  degree centrality & $HDC_2$=(389,378,522), $A_2$=(389,390,522) & 12 grid cells  & 0.93 \\
  \cline{2-4}
  region & $HDC_3$=(376,403,509), $A_3$=(392,403,509) & 16 grid cells  & 0.94 \\
  \hline
  Low & $LDC_1$=(375,396,520), $B_1$=(375,388,520) & 8 grid cells  & 0.58 \\
  \cline{2-4}
  degree centrality & $LDC_2$=(403,388,497), $B_2$=(403,388,509) & 12 grid cells  & 0.61 \\
  \cline{2-4}
  region & $LDC_3$=(374,390,521), $B_3$=(390,390,521) & 16 grid cells  & 0.49 \\
  \hline
\end{tabular}}
\end{table}

As mentioned in the Methods section, the turbulent energy field is fundamental to characterize the network. We checked a posteriori that from a localized information - such as the energy time series at a fixed point of the field - the network is able to infer the spatial behaviour of the surroundings, which involves velocity gradients, i.e., the vorticity field. Building the network from the vorticity field would have introduced spatial variations, by requiring a higher order information and leading to analogous results in terms of network.

\subsection{Sensitivity Analysis}

In the end, we performed a sensitivity analysis of the results regarding the reference sphere ($r=0.12$ and $C=(391,391,512)$). We first considered different thresholds, $\tau$, for the link activation. Then, another network based on a different reference sphere ($r=0.12$ and centered in $C'=(530,673,475)$) was studied.

\begin{figure}[h]
\includegraphics[width=\columnwidth]{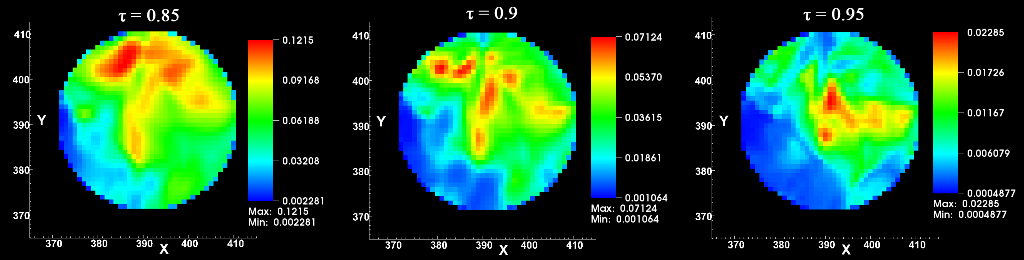}
\caption{Normalized degree centrality, $k_i$, on the $z=512$ plane. (left) $\tau=0.85$. (center) $\tau=0.9$. (right) $\tau=0.95$. Different color scales are adopted.}
\end{figure}

Beside $\tau=0.9$, networks for two different values were analyzed, $\tau=0.85$ and $\tau=0.95$. In Fig. 6 the normalized degree centrality on the plane $z=512$ is reported for the three $\tau$ values. In Table 2, some topological and spatial features of the three networks are compared. As $\tau$ decreased, the number of active links, $m$, increased at a faster rate than the size of the network (total number of nodes, $N$), while the cardinality $N_{int}$ does not change in any case. As a consequence, the degree centrality averagely increased and the high $k_i$ regions were more spatially expanded for decreasing $\tau$ values. The average physical distance, $L$, for the nodes $HDC=(385,401,508)$ and $LDC=(372,387,510)$ increases for decreasing $\tau$, similarly to what happens for the degree centrality. Despite the specific values assumed and the qualitative changes induced by the three threshold values, the spatial pattern detection is essentially independent from the choice of the threshold. Weighted networks, though computationally more expensive, can be adopted in future work as they can made results more robust against threshold variations.

\begin{table}[h]
\tbl{Topological and spatial features of the networks with $\tau=0.85, 0.9, 0.95$. $\bar{k} = \sum_{i=1}^N k_i/N$ is the mean degree centrality of the network, $\bar{L}=\sum_{i=1}^N L_i/N$ is the averaged physical distance computed as mean value of the network. $HDC=(385,401,508)$ and $LDC=(372,387,510)$.}
{\begin{tabular}{|c|c|c|c|}
  \hline
   \,\,\, \,\,\, & \,\,\, $\tau=0.85$ \,\,\, & \,\,\, $\tau=0.9$ \,\,\, & \,\,\, $\tau=0.95$ \,\,\, \\
   \hline
  \,\,\, Nodes \emph{N} \,\,\, & 172713 & 128785 & 75874 \\
   \hline
  \,\,\, Links \emph{m} \,\,\, & 243355115 & 80920781 & 11061400 \\
   \hline
  \,\,\, $\bar{k}$ \,\,\, & $1.632 \cdot 10^{-2}$ & $9.758 \cdot 10^{-3}$ & $3.843 \cdot 10^{-3}$ \\
   \hline
  \,\,\, $\bar{L}$ \,\,\, & $9.060 \cdot 10^{-2}$ & $7.703 \cdot 10^{-2}$ &  $4.621 \cdot 10^{-2}$ \\
   \hline
  \,\,\, $L_{HDC}$ \,\,\, & $8.377 \cdot 10^{-2}$ & $7.515 \cdot 10^{-2}$ & $3.784 \cdot 10^{-2}$ \\
   \hline
  \,\,\, $L_{LDC}$ \,\,\, & $5.931 \cdot 10^{-2}$ & $2.985 \cdot 10^{-2}$ & $1.146 \cdot 10^{-2}$ \\
  \hline
\end{tabular}}
\end{table}

A different spherical subdomain with radius $r=0.24$ centered in $C'=(530,673,475)$ was analyzed to build a new network with $\tau=0.9$ and the complete kinetic energy temporal series (1-1024). The physical distance between nodes $C$ and $C'$ is about 1.93, that largely exceeds the integral scale, $L = 1.376$. Nodes $C$ and $C'$ are far enough so that the two influence regions do not physically overlap. The new reference sphere has radius $r=0.12$ and center $C'=(530,673,475)$. This region has a spatio-temporal distribution for the kinetic energy which is different from the previous sphere centered in $C$ and this results into a new network having different cardinality and topology.

\begin{figure}[h]
\includegraphics[width=\columnwidth]{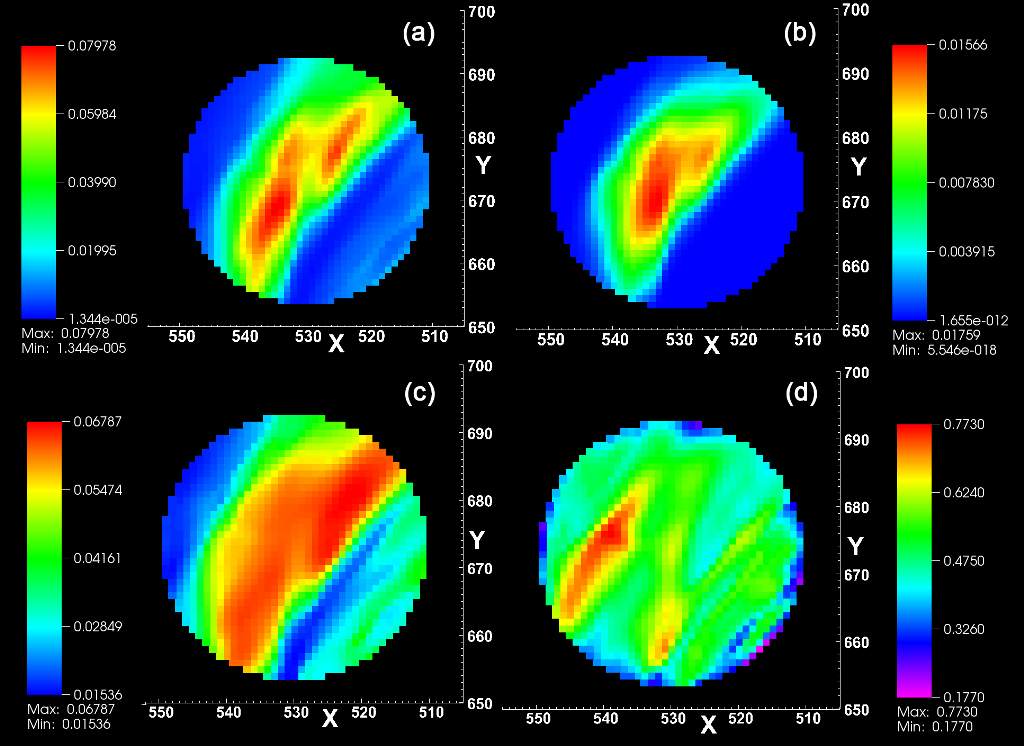}
\caption{Network centered in $C'=(530,673,475)$. (a) Normalized degree centrality, (b) eigenvector centrality, (c) average physical distance, (d) local clustering coefficient. Results are displayed on the section $z=475$.}
\end{figure}

\noindent In Fig. 7 the network results in terms of normalized degree centrality, eigenvector centrality, average physical distance, and local clustering coefficient are reported on a 2D section of the plane $z=475$. In Table 3, structural properties of the networks centered in $C$ and $C'$ are given for comparison. Despite the different shape assumed by the network metrics, results are of the same order of magnitude of those observed in the network centered in $C=(391,391,512)$.

\begin{table}[h]
\tbl{Topological features of the networks centered in $C=(391,391,512)$ and $C'=(530,673,475)$. $\bar{k} = \sum_{i=1}^N k_i/N$ is the mean degree centrality of the network, $\bar{L}=\sum_{i=1}^N L_i/N$ is the averaged physical distance computed as mean value of the network, $\bar{C} = \sum_{i=1}^N C_i/N$ is the global clustering coefficient.}
{\begin{tabular}{|c|c|c|}
  \hline
   \,\,\, \,\,\, & \,\,\, $C=(391,391,512)$ \,\,\, & \,\,\, $C'=(530,673,475)$ \,\,\, \\
   \hline
  \,\,\, Nodes \emph{N} \,\,\, & 128785 & 74432 \\
   \hline
  \,\,\, Links \emph{m} \,\,\, & 80920781 & 38799854 \\
  \hline
  \,\,\, $\bar{k}$ \,\,\, & $9.758 \cdot 10^{-3}$ & $1.400 \cdot 10^{-2}$ \\
  \hline
  \,\,\, $\bar{L}$ \,\,\, & $7.703 \cdot 10^{-2}$ & $4.853 \cdot 10^{-2}$ \\
  \hline
  \,\,\, $\bar{C}$ \,\,\, & $7.211 \cdot 10^{-1}$ & $6.969 \cdot 10^{-1}$ \\
  \hline
  \end{tabular}}
\end{table}

We can conclude that the spatial characterization is therefore independent of the chosen threshold, provided this latter is sufficiently high. Moreover, the presence of spatial patterns with different size and intensity is not limited to the chosen domain portion but can involve the whole turbulent field.

\section{Conclusions}

In the present work, the complex networks instruments were applied to analyze a forced isotropic turbulent field. Differently to recent literature studies which transformed each time series into a different network, here a single global network was built from spatio-temporal data following a two-point correlation approach carried out for all the pairs of selected nodes. The kinetic energy time series of the grid cells was chosen to define a monolayer network. A link between two nodes is active if the distance and statistical interdependence between two nodes are above suitably selected thresholds. Degree centrality, $k$, and average physical distance, $L$, were the best metrics able to quantify the spatial dynamics. High degree centrality regions evidenced spatial patterns coherently moving with similar vorticity over the large eddy turnover time scale. An indication of the spatial size of these regions was suggested by the average physical distance, varying from small scales up to the Taylor microscale.

\noindent The network analysis allowed us to handle big data and systematically identify different spatial regions. This goal would not have been so easily feasible without the use of the network metrics, which synthesized in a single framework a huge amount of detailed information. The proposed approach can suggest new insights into the spatial characterization of turbulent flows and, based on present findings, the application to highly inhomogeneous flows - such as compressible or wall flows - seems to be promising and is worth additional future investigation.

\bibliography{Scarsoglio}
\end{document}